\documentclass[twocolumn,twoside,preprintnumbers,amsmath,amssymb,showkeys]{revtex4}

\usepackage{epsfig}

\usepackage{graphicx}

\usepackage{fancyhdr}
\usepackage{pslatex}

\begin{document}
\title {Recent results of source function imaging from AGS through CERN SPS to RHIC}

\author{Roy A. Lacey}

\affiliation{Dept of Chemistry,
        SUNY Stony Brook, Stony Brook, NY 11794, USA
        }

\begin{abstract}

	Recent femtoscopic measurements involving the use of an 
imaging technique and a newly developed moment analysis are 
presented and discussed. We show that this new paradigm 
allows robust investigation of reaction dynamics for which the sound 
speed $c_s \neq 0$ during an extended hadronization period. 
Source functions extracted for charged pions 
produced in Au+Au and Pb+Pb collisions show non-Gaussian tails
for a broad selection of collision energies. The ratio of the RMS radii 
of these source functions in the out and side directions are found to 
be greater than 1, suggesting a finite emission time for pions.

\end{abstract}

\vskip -1.35cm

\maketitle

\thispagestyle{fancy}

\setcounter{page}{1}

\bigskip

\section{Introduction}

		Lattice calculations of the thermodynamical functions of quantum chromodynamics (QCD)
indicate a rapid transition from a confined hadronic phase to a chirally symmetric de-confined 
quark gluon plasma (QGP), for a critical temperature 
$T_c \sim 170$~MeV \cite{Karsch:2000ps}. 
The creation and study of this QGP is one of the most important goals 
of current ultra-relativistic heavy ion research \cite{qgp05}.
Central to this endevour, is the question of whether or not the QGP phase 
transition is first order ($\Delta T = 0$), $2^{\text{nd}}$ order, or a 
rapid crossover reflecting an increase in the entropy density associated with the change 
from hadronic ($d_H$) to quark and gluon ($d_Q$) degrees of freedom. 
Currently, lattice calculations seem to favor a rapid crossover \cite{Karsch:2006sm,Karsch:2000kv}
i.e. $\Delta T \ne 0$ for the transition.

	An emitting system which undergoes a sharp first order QGP phase transition is 
predicted to show a larger space-time extent, than that for a purely 
hadronic system \cite{Pratt:1984su,Rischke:1996em}.
Here, the rationale is that the transition to the QGP ``softens" the equation 
of state (ie. the sound speed $c_s \sim 0$) in the transition region, and 
this delays the expansion and considerably prolongs the lifetime of the system. 
That is, matter-flow through this "soft region" of energy densities
during the expansion phase, will temporarily slow down and could even 
stall under suitable conditions.

\begin{figure}[!htb]
\begin{center}
\vskip0.15cm
\includegraphics[width=1.0\linewidth]{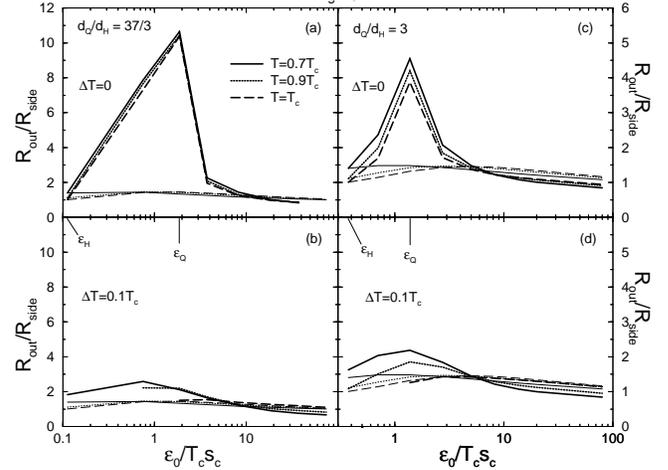} 
\end{center}
\vskip-0.5cm
\caption{\emph{ \small $R_{out}/R_{side}$ as a function of the initial 
energy density (in units of $T_cs_c$) for an expanding fireball. 
(a,b) are for $d_Q/d_H = 37/3$, (c,d) for $d_Q/d_H = 3$. 
The thick lines in (a,c) are for $\Delta T = 0$, in (b,d) for 
$\Delta T = 0.1T_c$. Thin lines show results for an ideal gas case. 
Solid lines show results for $T = 0.7T_c$, dotted for $T = 0.9T_c$, 
dashed for $T = T_c$. The figure is taken from Ref.~\cite{Rischke:1996em}.
}}
\label{Fig1_Rischke_RoutRside}
\end{figure}

	To explore this prolonged lifetime, it has been common 
practice to measure the widths of emission source 
functions (assumed to be Gaussian) in the out- side- and 
long-direction ($R_{out}$, $R_{side}$ and $R_{long}$)
of the Bertsch-Pratt coordinate system \cite{Lisa:2005dd,Adler:2001zd,
Adcox:2002uc,Adams:2003vd}. For such extractions, the Coulomb effects 
on the correlation function are usually assumed to be 
separable \cite{Sinyukov:1998fc} as well.
The ratio $R_{out}/R_{side} > 1$, is expected  
for systems which undergoe a strong first order QGP phase 
transition \cite{Pratt:1984su,Rischke:1996em}. 
This is clearly illustrated in Fig.~\ref{Fig1_Rischke_RoutRside}, 
where the values obtained from a hydrodynamical model 
calculation \cite{Rischke:1996em} are plotted as a function of 
energy density (in units of $T_cs_c$; $s$ is the entropy density). 
These rather large ratios (especially those in Fig.~\ref{Fig1_Rischke_RoutRside}a 
and c) have served as a major motivating factor for experimental 
searches of a prolonged lifetime at several accelerator 
facilities \cite{Lisa:2005dd,Ganz:1998zj,Bearden:2003ku,Adler:2004rq,
Adler:2001zd,Adcox:2002uc,Adams:2003vd}.

	An interesting aspect of Fig.~\ref{Fig1_Rischke_RoutRside} 
is the strong dependence of $R_{out}/R_{side}$ on the width of the 
transition region (i.e. $\Delta T \ne 0$).  
A comparison of Figs.~\ref{Fig1_Rischke_RoutRside}a and b shows that 
this ratio 
is considerably reduced (by as much as a factor of four)
when the calculations are performed for $\Delta T = 0.1T_c$. 
Thus, the space-time extent of the emitting 
system is rather sensitive to the reaction dynamics and whether or 
not the transition is a cross over. 
It has also been suggested that the shape of the emission source function can 
provide signals for a second order phase transition and whether or not particle 
emission occurs near to the critical end point in the QCD 
phase diagram \cite{Csorgo:2005it}. 

\section{Reaction dynamics at RHIC}

		Experimental evidence for the creation of locally equilibrated 
nuclear matter at unprecedented energy densities at RHIC, is 
strong \cite{Adcox:2004mh,Adams:2005dq,Back:2004je,Arsene:2004fa,
Gyulassy:2004zy,Muller:2004kk,Shuryak:2004cy,Heinz:2001xi}.
Jet suppression studies indicate that the constituents of this matter 
interact with unexpected strength and this matter is almost opaque to 
high energy partons \cite{Adler:2002tq,Adler:2005ee}. 
\begin{widetext}
%

 \begin{figure}[tb]
 \includegraphics[width=0.75\linewidth]{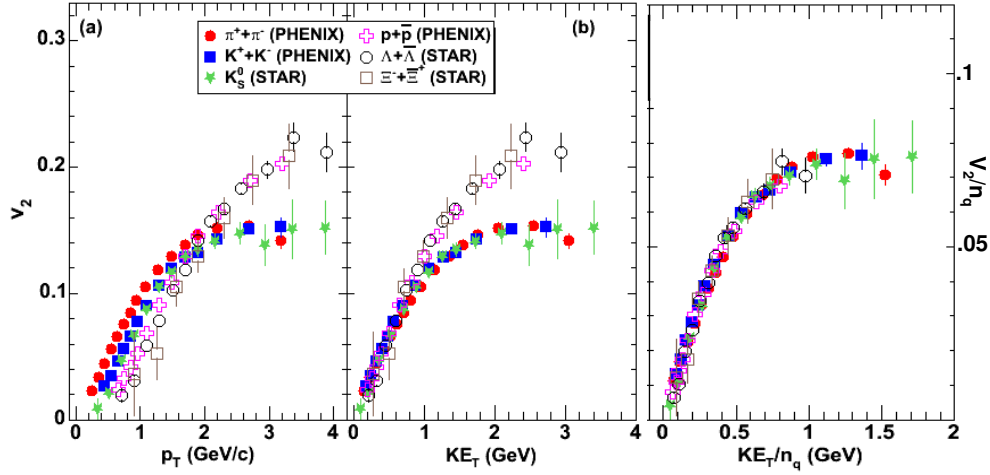}  
 \caption{\label{v2KETn}
	(Color online) $v_2$ vs. $p_T$ (left panel) and $KE_T$ (middle panel). The 
scaled results in the right panel is obtained via $n_q$ scaling of the data 
shown in the middle panel. Results are shown for several particle species 
produced in minimum bias Au+Au collisions at $\sqrt{s_{NN}} = 200$ 
GeV \cite{Adare:2006ti} 
}
\end{figure}
\end{widetext}
Elliptic flow measurements \cite{Issah:2006qn,Adare:2006ti} validate the 
predictions of perfect fluid hydrodynamics for the scaling of the elliptic 
flow coefficient $v_2$ with eccentricity $\varepsilon$, system size and 
transverse kinetic energy KE$_T$ \cite{Csanad:2005gv,Csanad:2006sp,Bhalerao:2005mm}; 
they also indicate the predictions of valence quark number ($n_q$)
scaling \cite{Voloshin:2002wa,Fries:2003kq,Greco:2003mm}.
The result of such scaling is illustrated in Fig.~\ref{v2KETn}c; it shows
that, when plotted as a function of the transverse kinetic energy KE$_T$ and 
scaled by the number of valence quarks $n_q$ of    
a hadron ($n_q = 2$ for mesons and $n_q = 3$ for baryons), $v_2$ shows a universal 
dependence for a broad range of particle 
species \cite{Issah:2006qn,Adare:2006ti,Lacey:2006pn}. This has been
interpreted as evidence that hydrodynamic expansion of the QGP occurs 
during a phase characterized by (i) a rather low 
viscosity to entropy ratio $\eta/s$ \cite{Shuryak:2004cy,Gyulassy:2004zy,
Heinz:2001xi,Asakawa:2006tc,Lacey:2006pn} and (ii) independent quasi-particles which exhibit 
the quantum numbers of quarks \cite{Voloshin:2002wa,Fries:2003kq,Greco:2003mm,
Xu:2005jt,Muller:2006ee,Lacey:2006pn}. 

	The scaled $v_2$ values shown in Fig. \ref{v2KETn}c allow an estimate 
of $c_s$ because the magnitude of $v_2/\epsilon$ 
depends on the sound speed \cite{Bhalerao:2005mm}.  
One such estimate \cite{Issah:2006qn,Adare:2006ti} gives $c_s \sim 0.35 \pm 0.05$; a value 
which suggests an effective equation of state (EOS) which is softer than that for the high 
temperature QGP \cite{Karsch:2006sm}. It however, does not reflect a strong first order phase 
transition in which $c_s \sim 0$ during an extended hadronization period. This 
sound speed is also compatible with the fact that $v_2(p_T)$ is observed to saturate in Au+Au 
collisions for the collision energy range $\sqrt{s_{NN}} = 60 - 200$ GeV \cite{Adler:2004cj}.

	Femtoscopic measurements involving the use of the 
Bowler-Sinyukov 3D HBT analysis method [in Bertsch-Pratt 
coordinates], have been used to probe for long-range emissions 
from a possible long-lived source \cite{Lisa:2005dd,Ganz:1998zj,
Bearden:2003ku,Adler:2004rq,Adler:2001zd,Adcox:2002uc,Adams:2003vd}.
The observed RMS-widths for each dimension of the emission source 
$R_{\text{long}}, R_{\text{side}}$ and $R_{\text{out}}$, 
show no evidence for such emissions. That is, $R_{\text{out}}/R_{\text{side}} \sim 1.0$. 
It is somewhat paradoxical that these observations have been interpreted 
as a femtoscopic puzzle \cite{Lisa:2005dd}
-- ``the HBT puzzle" -- despite the fact that they clearly reflect a 
rich and complex set of thermodynamic trajectories for 
which $c_s \neq 0$ (or $\Delta T \neq 0$) during an extended 
hadronization period. 

%
%
%
 \begin{figure}[tb]
 \begin{center}
 \includegraphics[width=0.75\linewidth]{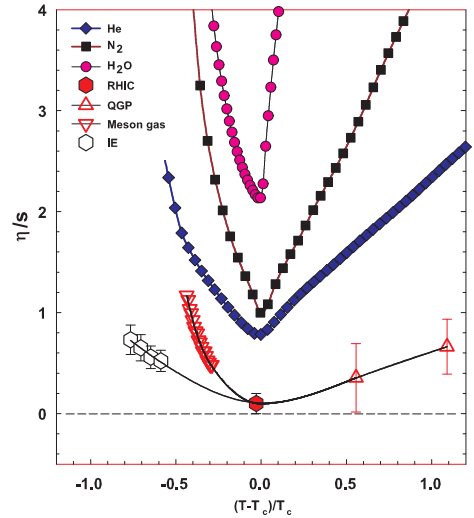}  
 \caption{\label{eta_over_s}
	(Color online) $\eta/s$ vs $(T-T_c)/T_c$ for several substances as indicated.
	The calculated values for the meson-gas have an associated error 
	of $\sim$ 50\%~\cite{Chen:2006ig}. 
	The lattice QCD value $T_c = 170$~MeV~\cite{Karsch:2000kv} 
	is assumed for nuclear matter. The lines are drawn to guide the eye.
}
\end{center}
 \end{figure}

	Hints for even more complicated reaction dynamics is further illustrated 
via a comparison of $\eta/s$ values for nuclear,
atomic and molecular substances in Fig.~\ref{eta_over_s}. 
It shows the observation that for atomic and molecular substances, 
the ratio $\eta/s$ exhibits a minimum of comparable depth 
for isobars passing in the vicinity of the liquid-gas
critical point \cite{Kovtun:2004de,Csernai:2006zz}. When an isobar passes 
through the critical point (as in Fig.~\ref{eta_over_s}), the minimum 
forms a cusp at $T_c$; when it passes below the critical point, the minimum 
is found at a temperature below $T_c$ (liquid side) but is accompanied by a 
discontinuous change across the phase transition. For an isobar passing 
above the critical point, a less pronounced minimum is found at a value 
slightly above $T_c$. The value $\eta/s$ is smallest in the vicinity of $T_c$ 
because this corresponds to the most difficult condition for the transport of 
momentum \cite{Csernai:2006zz}.

	Given these observations, one expects a broad range of trajectories in 
the $(T,\mu_B)$ plane for nuclear matter, to show $\eta/s$ minima 
with a possible cusp at the critical point ($\mu_B$ is the baryon chemical 
potential). The exact location of this point is of course not known, and only coarse 
estimates of where it might lie are available. The open triangles in 
the figure show calculated values 
for $\eta/s$ along the $\mu_B=0$, $n_B=0$ trajectory. For $T < T_c$ the $\eta/s$ 
values for the meson-gas show an increase for decreasing values of $T$. 
For $T$ greater than $T_c$, the lattice results \cite{Nakamura:2004sy} 
indicate an increase of $\eta/s$ with $T$, albeit with large error bars.

	These trends suggest a minimum for $\eta/s$, in the 
vicinity of $T_c$, whose value is close to the conjectured absolute lower bound 
$\eta/s = 1/4\pi$. Consequently, it is tempting to speculate that 
it is the minimum expected when the hot and dense QCD matter 
produced in RHIC collisions follow decay trajectories which are close to 
the QCD critical end point (CEP). Such trajectories could be readily 
followed if the CEP acts as an attractor for thermodynamic trajectories of 
the decaying matter \cite{Nonaka:2004pg,Kampfer:2005nt}. 

	The above insights on the reaction dynamics of hot and dense QCD 
matter, strongly suggest that full utility of the femtoscopic probe 
requires detailed measurements and theoretical investigations 
of both the shape and the space-time character of emission source functions.	
Here, it should be stressed that a good understanding of the space-time 
evolution of the QGP is crucial. 
This is because, in contrast to other signals, it is derived from the 
influence of the equation of state on the collective dynamical evolution 
of the system. This is less clear-cut for many other proposed signals
\cite{Kapusta:1991qp,Gerschel:1992gy}.

\section{Source imaging: A femtoscopic paradigm shift}
\subsection{1D source functions via imaging}

	An initial strategy has been to extract the ``full" distribution  
of the 1D emission source function \cite{Adler:2006as}. The 1D~Koonin-Pratt 
equation \cite{Koonin:1977fh}:
%
%
\begin{equation}
  C(q)-1 = 4\pi\int dr r^2 K_0(q,r) S(r),
%
%
\label{kpeqn}
\end{equation}
relates this source function $S(r)$ (i.e. the probability of emitting 
a pair of particles at a separation $r$ in the pair center of 
mass (PCMS) frame) to the 1D correlation function $C(q)$.
The angle-averaged kernel $K_0(q,r)$ encodes the the final state interaction 
(FSI) which is given in terms of the final state wave function 
$\Phi_{\bf q}({\bf r})$, as $K_0(q,r)=\frac{1}{2}\int
d(\cos(\theta_{\bf q, r})) (|\Phi_{\bf q}({\bf r})|^2-1)$, where $\theta_{\bf q, r}$ 
is the angle between {\bf q} and {\bf r}~\cite{Brown:2000aj}. 
%
\begin{figure}[tbh]
\includegraphics[width=0.90\linewidth]{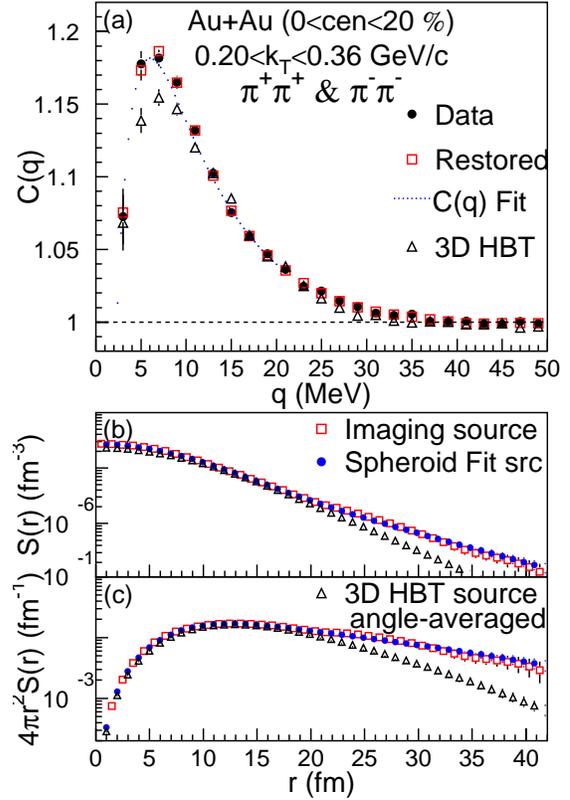}
\caption{\label{fig1} (color online)
(upper, a) (filled circles) Correlation function, C(q) 
for $\pi^+\pi^+$ and $\pi^-\pi^-$ pairs;  
(open squares) restored correlation from imaging technique; 
(dotted line) direct correlation fitting; 
(open triangles) 1D angle-averaged correlation of 3D correlation function.  
(lower) 1D source function (b) S(r) and (c) 4$\pi r^2$S(r): 
(open squares) imaging; 
(filled circles) spheroid fit to correlation function; 
(open triangles) angle-averaging of 3D-Gaussian source function. 
The figure is reproduced from Ref. \cite{Adler:2006as}
}
\end{figure}
%

	 The model-independent imaging technique of Brown and 
Danielewicz~\cite{Brown:1997ku,Brown:2000aj} is used to 
invert Eq.~\ref{kpeqn} to obtain $S(r)$. In brief, it utilizes a numerical 
calculation of the two particle wave function (including FSI) to produce 
an inversion matrix that operates on the measured correlation function 
to give the associated source function. A detailed description of the 
procedure for robust measurement of the correlation functions is 
given in these proceedings \cite{pchung_Ismd}.

	The open squares in Fig.~\ref{fig1}(b) show the 1D source image
obtained from the correlation function (filled circles) presented 
in Fig.~\ref{fig1}(a). The open squares in Fig.~\ref{fig1}(a) show the 
corresponding restored correlation function obtained via Eq.~\ref{kpeqn} with the 
extracted source function as input. Their consistency serves as a 
good cross check of the imaging procedure.
	
	  Figure \ref{fig1}(b)) indicates a Gaussian-like pattern at 
small $r$ and an unresolved ``tail" at large $r$. For comparison,  
the source function constructed from the parameters 
($R_{\text{long}}, R_{\text{side}}$, $R_{\text{out}}$ and $\lambda$), 
obtained in an earlier 3D HBT analysis \cite{Adler:2004rq} is 
also shown; the associated 3D angle-averaged correlation 
function is shown in Fig.~\ref{fig1}(a).
The imaged source function exhibits a more prominent 
tail than the angle averaged 3D HBT source function. 
These differences reflect the differences in the 
associated correlation functions for $q \alt 10~-~15$~MeV,
as shown in Fig. \ref{fig1}(a).
This could come from the Gaussian shape assumption 
employed for  $S(r)$ in the 3D HBT analysis. That is, 
the 3D Gaussian fitting procedure by construction is sensitive 
only to the main component of $S(r)$, and thus would not be capable 
of resolving fine structure at small-q/large-r.

	The radial probabilities (4$\pi r^2$S(r)) are also compared
in Fig.\ref{fig1}(c); they show that the differences are actually 
quite large. They are clearly related to the long-range contribution 
to the emission source function.

\subsection{1D source function via correlation function fitting}

	  Another approach for 1D source function extraction is to 
perform direct fits to the correlation function.
This involves the determination of a set of values for an assumed 
source function shape, which reproduce the observed correlation 
function when the resulting source function is 
inserted into Eq.~\ref{kpeqn}.  
The filled circles in Figs.~\ref{fig1}(b,c) show the source 
function obtained from such a fit for a Spheroidal shape~\cite{Brown:2000aj}, 
%
%
\begin{equation}
S(r) = \frac{\lambda \: R_{\text{eff}} \times e^{ - \frac {r^2}{4R_T^2} } 
\text{erfi}(\frac{r}{2 R_{\text{eff}} }) }{\left( \text{8} \pi R_T^2 R_0 r \right)}, \; \text{for} \; R_0 > R_T, 
%
\label{blimp}
\end{equation}
%
%
where $R_{\text{eff}} = 1/ \sqrt{(1/{R_T^2} - 1/{R_0^2}) }$, $R_{T}$ is the 
radius of the Spheroid in two perpendicular spatial dimensions and 
$R_0 = a \times R_T$ is the radius in the third spatial dimension; 
$a$ is a scale factor.  
The long axis of the Spheroid is assumed to be oriented in the out 
direction of the Bertsch-Pratt coordinate system.
The fraction of pion pairs which contribute to the source $\lambda$,
is given by the integral of the normalized source function over the 
full range of $r$. 
The source function shown in Figs.~\ref{fig1}(b) and \ref{fig1}(c), 
indicates remarkably good agreement with that obtained via the 
imaging technique, confirming a long-range contribution to the 
emission source function.  

\subsection{Origin of the long-range contributions to the 1D source function}

	Extraction of the detailed shape of the 1D source function is made 
ambiguous by the observation that several assumed shapes gave equally 
good fits which reproduced the observed correlation function. 
Nonetheless, one can ask whether or not a simple kinematic transformation 
from the longitudinal co-moving system (LCMS) to the PCMS can account
for the observed tail?  Here, the point is that instantaneous freeze-out 
of a source with $R_{out}/R_{side} \sim 1$ in the LCMS would give a maximum 
kinematic boost so that $R_{out} \sim \gamma\times R_{side}$ in the PCMS
($\gamma$ is the Lorentz factor). 
Another possibility could be that the particle emission source is 
comprised of a central core and a halo of long-lived resonances. 
For such emissions, the pairing between particles from the core and 
secondary particles from the halo is expected to dominate the 
long-range emissions \cite{Nickerson:1997js} and possibly give rise to 
a tail in the source function.

	Both of these possibilities have been checked \cite{Adler:2006as}; the 
indications are that the observations can not be explained solely by either 
one. However, further detailed 3D imaging measurements were deemed necessary
to quantitatively pin down the origin of the tail in the source function and 
to determine its possible relationship to the reaction dynamics.

\section{3D source Imaging}

	In order to develop a clearer picture of the reaction dynamics across 
beam energies, an ambitious [ongoing] program was developed to measure 3D source 
functions in heavy ion collisions from AGS through CERN SPS to RHIC energies. 
This program exploits the cartesian harmonic decomposition technique of 
Danielewicz and Pratt \cite{Danielewicz:2005qh,Chung:2006rz} in conjunction with imaging 
and fitting, to extract the pair separation distributions in the out-, side- and long-directions.
Model comparisons to these distributions are expected to give 
invaluable insights on the reaction dynamics.

\subsection{Cartesian harmonic decomposition of the 3D correlation function}

Following Danielewicz and Pratt  \cite{Danielewicz:2005qh}, the 3D correlation function 
can be expressed in terms of correlation moments as:
\begin{equation}
C(\mathbf{q}) - 1 = R(\mathbf{q}) = \sum_l \sum_{\alpha_1 \ldots \alpha_l}
   R^l_{\alpha_1 \ldots \alpha_l}(q) \,A^l_{\alpha_1 \ldots \alpha_l} 
   (\Omega_\mathbf{q}),
\label{eqn3}
\end{equation}
where $l=0,1,2,\ldots$, $\alpha_i=x, y, z$, 
$A^l_{\alpha_1 \ldots \alpha_l}(\Omega_\mathbf{q})$
are cartesian harmonic basis elements ($\Omega_\mathbf{q}$ is solid angle 
in $\mathbf{q}$ space) and $R^l_{\alpha_1 \ldots \alpha_l}(q)$ are 
cartesian correlation moments given by
\begin{equation}
 R^l_{\alpha_1 \ldots \alpha_l}(q) = \frac{(2l+1)!!}{l!}
 \int \frac{d \Omega_\mathbf{q}}{4\pi} A^l_{\alpha_1 \ldots \alpha_l} (
 \Omega_\mathbf{q}) \, R(\mathbf{q}).
\label{eqn4}
\end{equation}
In Eqs.~\ref{eqn3} and \ref{eqn4}, the coordinate axes are oriented so that z 
is parallel to the beam (i.e. the long direction), x is orthogonal to z and points
along the total momentum of the pair in the LCMS frame (the out direction), 
and y points in the side direction orthogonal to both x and z.
 
	The correlation moments, for each order $l$, can be calculated from the 
measured 3D correlation function with Eq. \ref{eqn4}. 
Another approach is to truncate Eq.~\ref{eqn3} so that it includes 
only non-vanishing moments and hence, can be expressed in terms of 
independent moments. For instance,  
up to order $l=4$, there are 6 independent moments: 
$R^0$, $R^2_{x2}$, $R^2_{y2}$, $R^4_{x4}$, $R^4_{y4}$ and $R^4_{x2y2}$. 
Here, $R^2_{x2}$, $R^2_{y2}$, etc is used as a shorthand notation 
for $R^2_{xx}$, $R^2_{yy}$, etc.  
The independent moments can then be extracted as a function of $q$ by 
fitting the truncated series to the experimental 3D correlation function 
with the moments as the parameters of the fit. Moments have been  
so obtained up to order $l=8$, depending on available statistics. The 
higher order moments ($l\ge 4$) are generally found to be small or zero
especially for the lower collision beam energies.     

\begin{figure}[htb]
\begin{center}
\includegraphics*[angle=0, width=0.85\linewidth]{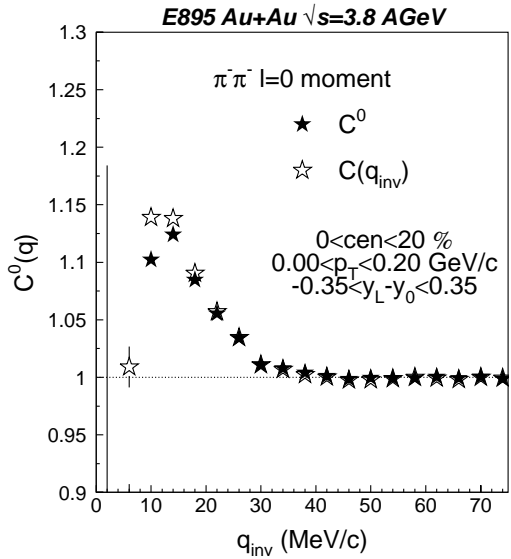}
\end{center}
\caption{\emph{ \small 1D correlation function C(q) (open stars) and $l = 0$ 
moment $C(q)^0=R^0(q)+1$ (solid stars) for low $p_T$ mid-rapidity $\pi^-\pi^-$ pairs 
from central Au+Au collisions at $\sqrt s = 3.8$ GeV. 
}}
\label{e895_leq0}
\end{figure}

\begin{figure}[htb]
\begin{center}
\includegraphics*[angle=0, width=0.95\linewidth]{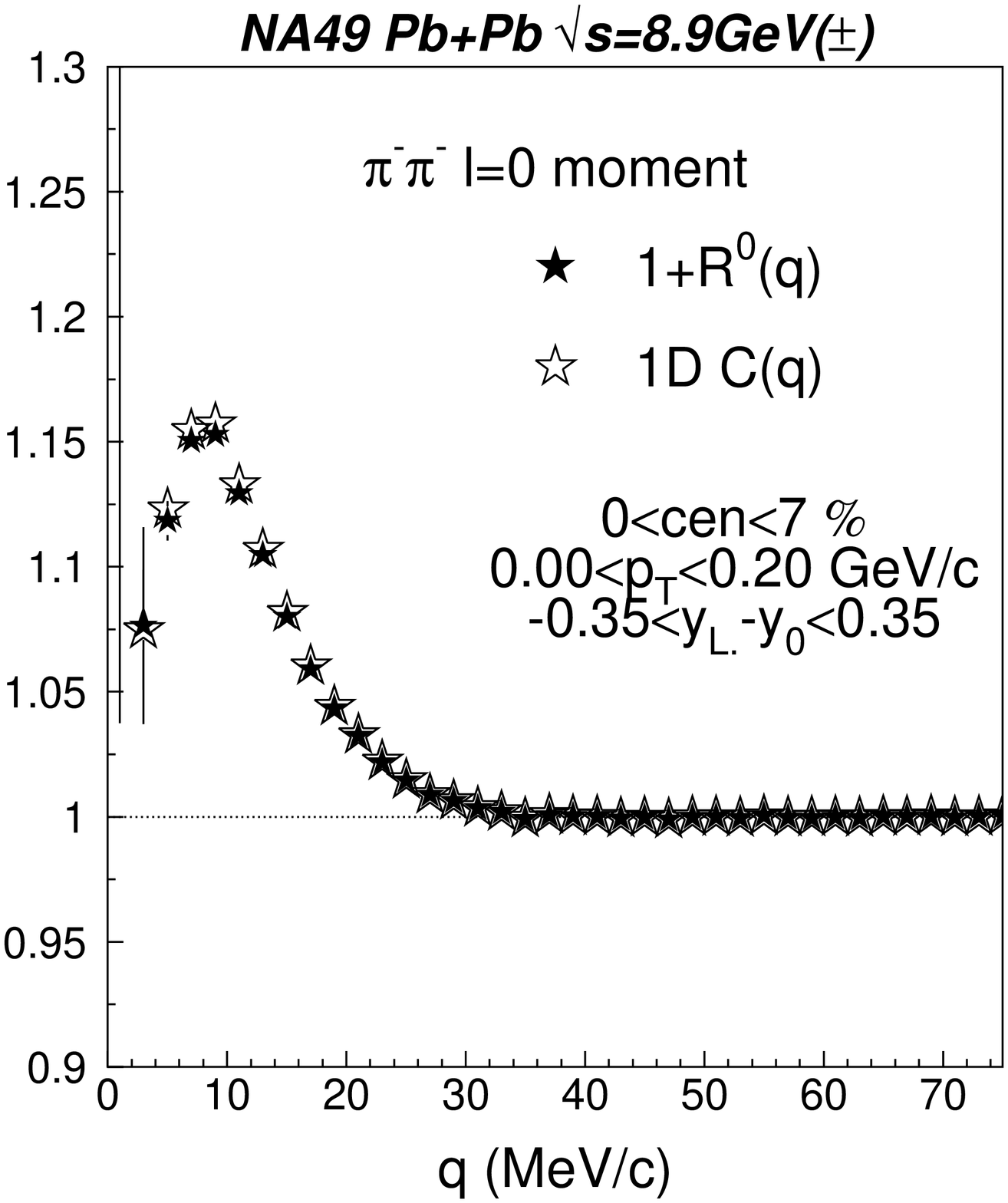}
\end{center}
\vskip-1.5cm
\caption{\emph{ \small 1D correlation function C(q) (open stars) and $l = 0$ 
moment $C(q)^0=R^0(q)+1$  (solid stars) for low $p_T$ mid-rapidity $\pi^-\pi^-$ pairs 
from central Pb+Pb collisions at $\sqrt s = 8.9$ GeV. 
}}
\label{40gevl0}
\end{figure}

\begin{figure}[htb]
\begin{center}
\includegraphics*[angle=0, width=0.95\linewidth]{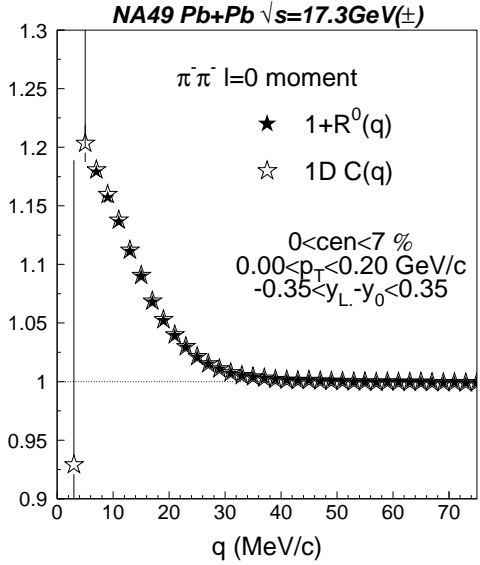}
\end{center}
\vskip-1.5cm
\caption{\emph{ \small Same as Fig.\ref{40gevl0} but for $\sqrt s = 17.3$ GeV. 
Result are averaged over runs with +ve and -ve magnetic fields.
}}
\label{160gevl0}
\end{figure}

\begin{figure}[htb]
\begin{center}
\includegraphics*[angle=0, width=0.85\linewidth]{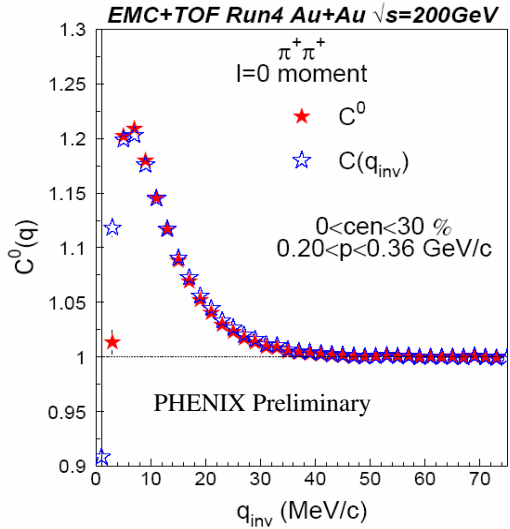}
\end{center}
\caption{\emph{ \small 1D correlation function C(q) (open stars) and $l = 0$ 
moment $C(q)^0=R^0(q)+1$  (solid stars) for low $p_T$ mid-rapidity $\pi^-\pi^-$ pairs 
from central Au+Au collisions at $\sqrt s = 200$ GeV. 
}}
\label{emc_tof_l0}
\end{figure}

	Figures \ref{e895_leq0} - \ref{emc_tof_l0} show representative comparisons 
between the 1D correlation functions, $C(q)$, and the $l = 0$ moment $C(q)^0 \equiv R^0(q)+1$ 
for mid-rapidity particles emitted in Au+Au and Pb+Pb collisions. 
For the AGS and SPS measurements, $|y_L-y_0|<0.35$, 
where $y_L$ is particle laboratory rapidity and $y_0$ is CM rapidity. For 
RHIC measurements $|y|<0.35$. 
The $p_T$ and centrality selections are as indicated in each figure.  
The $l = 0$ moments are in good agreement with the 1D correlation 
functions; such agreement is expected in the absence of significant pathology,
especially those related to the angular acceptance. Therefore, they attest to 
the reliability of the technique used for their extraction. 
Similarly good agreement was found for all other beam energies 
and analysis cuts.
%
%
%
%
%

\begin{figure}[hbt2]
\resizebox{19pc}{!}{\includegraphics{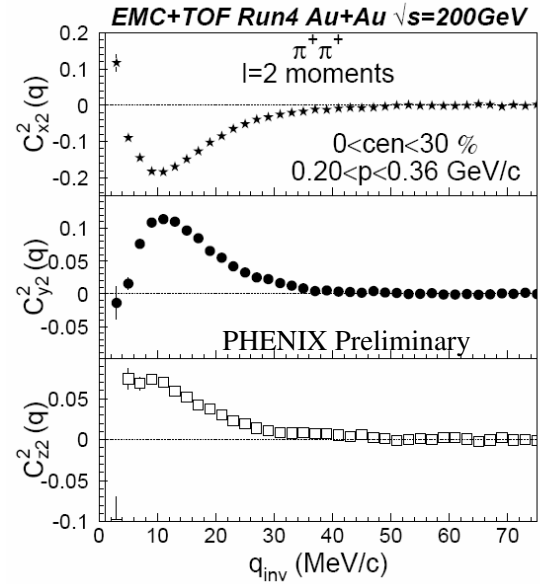}}
\resizebox{19pc}{!}{\includegraphics{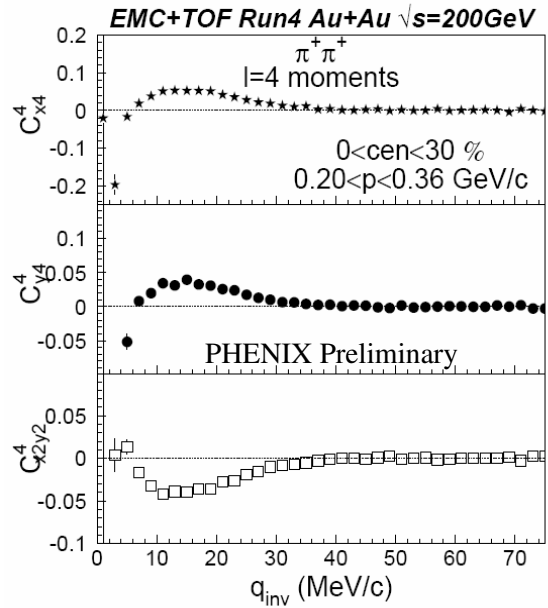}}
\caption{\emph{\small $l = 2$ (top panels) and $l = 4$ (bottom panels)
correlation moments for mid-rapidity (low $p_T$) $\pi^-\pi^-$ pairs 
from 0-30\% central Au+Au collisions at $\sqrt s = 200$ GeV. Note 
the change in notation; 
$C^l_{\alpha_1 \ldots \alpha_l}(q) \equiv R^l_{\alpha_1 \ldots \alpha_l}(q)$
}}
\label{emc_tof_l2l4}
\end{figure}

	Correlation moments for $l=2$ and $l=4$ are shown for Au+Au collisions 
obtained at RHIC, in Fig. \ref{emc_tof_l2l4}.  
The $l=2$ moments show an anti-correlation for $C^2_{x2}(q) \equiv R^2_{x2}(q)$ and 
positive correlations for $C^2_{y2}(q)$ and $C^2_{z2}(q)$, indicating that 
the emitting source is more extended in the out (x) direction and less so 
in the side (y) and long (z) directions, when compared to the angle 
average $C^0(q)$ (also see discussion on imaged moments). 
It is noteworthy that only two of these $l=2$ correlation 
moments are actually independent, 
i.e. $C^2_{x2}(q)+ C^2_{y2}(q)+ C^2_{z2}(q)=0$.

The bottom panel of Fig. \ref{emc_tof_l2l4} shows that rather sizable 
signals are obtained for $l=4$. Here, it is $C^4_{x2y2}$ correlation 
moment which shows an anti-correlation. Moments were also 
obtained for $l=6$. Since they were all found to be small, they do not 
have a significant influence on the shape of the source function. 

\begin{figure}[hbt2]
\resizebox{20pc}{!}{\includegraphics{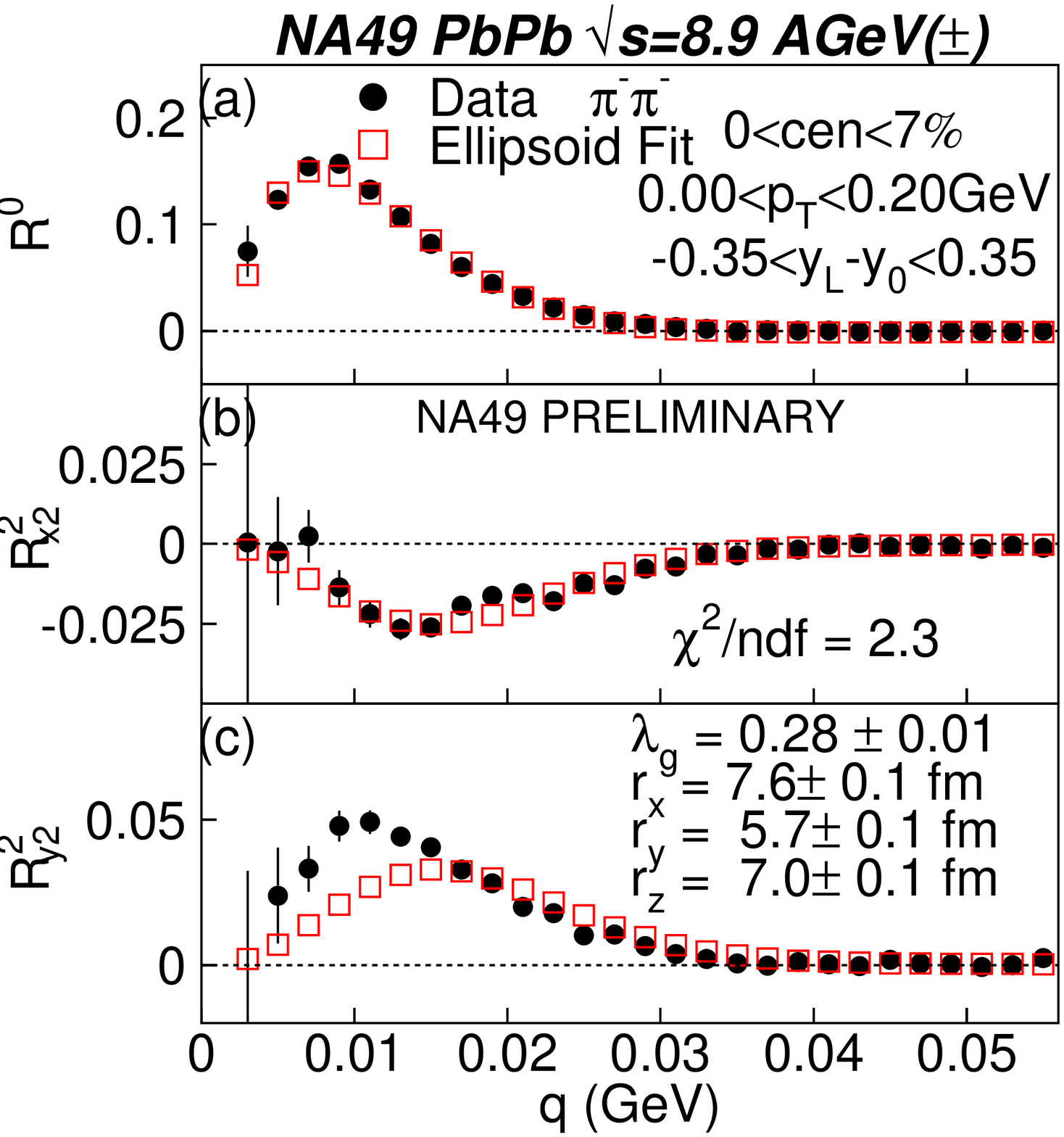}}
\vskip-2.0cm
\resizebox{20pc}{!}{\includegraphics{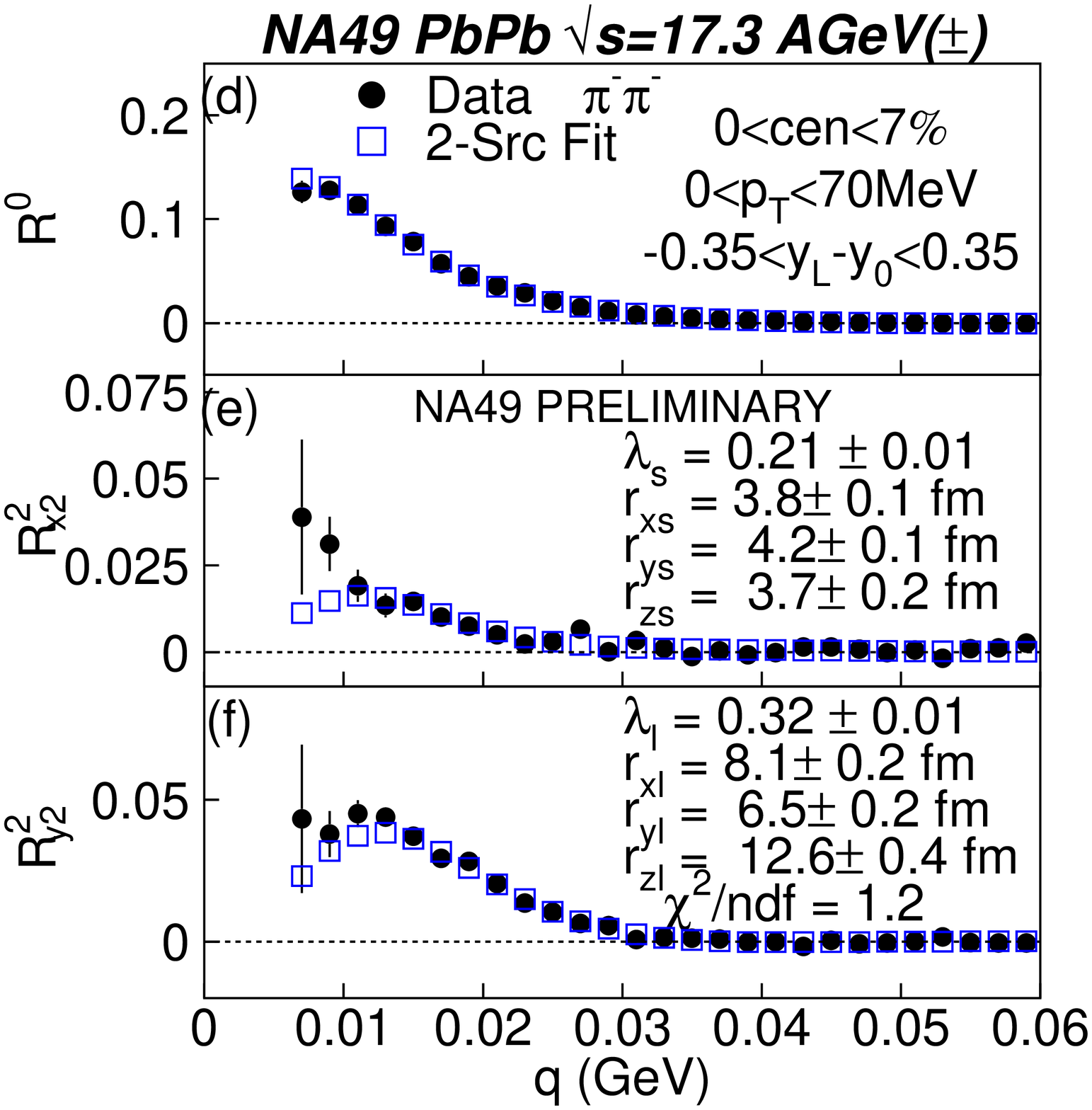}}
\vskip-1.5cm
\caption{\emph{\small $l = 2$ correlation moments for mid-rapidity, low $p_T$ $\pi^-\pi^-$ pairs 
from central Pb+Pb collisions for $\sqrt s = 8.9$ GeV (top panels) and 
$\sqrt s = 17.3$ GeV. The open squares represent the result of a simultaneous fit 
of the moments with an an assumed shape for the 3D source function (see text).
}}
\label{40gevfit}
\end{figure}

 	Figure \ref{40gevfit} shows correlation moments for Pb+Pb collisions 
obtained at SPS energies. The top and bottom panels show results for
$\sqrt s = 8.9$ and 17.3 GeV respectively. In each case, the rapidity 
and centrality selections are the same, but the $p_T$ selections are 
different as indicated. The open squares in the figure  represent 
the result of a simultaneous fit to the independent moments with  
assumed shapes for the source function as discussed below. Within errors,
the magnitudes of the higher order moments were found to be insignificant.
%
%
%

		The $R^2_{x2}(q)$ and $R^2_{y2}(q)$ correlation moments in the top panels of 
Fig. \ref{40gevfit} indicate an emission source which is more extended in the 
out and long directions (compared to $R^0(q)$). By contrast, the $R^2_{x2}$ 
and $R^2_{y2}(q)$ correlation moments in the bottom panels indicate positive 
correlations and hence an emission source with big extension (compared to $R^0(q)$) 
in the long direction. In part, this difference is related to a different 
kinematic factor for the two $p_T$ selections.

	The correlation moments shown in Figs. \ref{emc_tof_l2l4} and \ref{40gevfit} 
provide the basis for quantitative extraction of the pair separation distributions
in the out-, side- and long-directions.  
This is achieved either by fitting the moments or by imaging them. 

\subsection{Moment Fitting}

	As discussed earlier, the 3D correlation function can be represented 
as a sum of independent cartesian moments. Thus, a straight forward procedure 
for source function extraction is to assume a shape and perform a simultaneous 
fit to all of the measured independent moments. The open squares in the top 
panel of Fig. \ref{40gevfit} show the results of such a fit with an assumption
of a 3D Gaussian source function (termed ellipsoid since there is no 
requirement that all three radii are the same). The fit parameters are 
indicated in panel (c). Overall, the moments are reasonably well represented 
by an ellipsoid for $\sqrt s = 8.9$ GeV. However, a clear deviation is 
apparent for $R^2_{y2}$. A similar fit performed for the moments obtained 
for $\sqrt s = 17.3$ GeV show even larger non Gaussian deviations.   

	The bottom panels of Fig. \ref{40gevfit} show the results of a fit which 
assumes a linear combination of two Gaussians for the source function. The fit
parameters are indicated in the figure. The two-source shape yields a much 
better representation of the data for $\sqrt s = 17.3$ GeV.
The parameters obtained from these and the earlier fits, can be used 
to generate source functions in the out-, side- and long-directions 
for comparison with models.

\subsection{Moment Imaging}

	A complimentary approach for obtaining the source function 
distributions is that of imaging. In this case, both the 3D correlation 
function C($\mathbf{q}$) and source function S($\mathbf{r}$) are expanded 
in a series of cartesian harmonic basis with correlation moments 
$R^l_{\alpha_1 \ldots \alpha_l}(q)$ and source moments 
$S^l_{\alpha_1 \ldots \alpha_l}(q)$ respectively. Substitution into 
the 3D Koonin-Pratt equation
\begin{equation}
  C(\mathbf{q})-1 = \int d\mathbf{r} K(\mathbf{q},\mathbf{r}) S(\mathbf{r}),
  \label{3dkpeqn}
\end{equation}
gives \cite{Danielewicz:2005qh};
\begin{equation}
  R^l_{\alpha_1 \ldots \alpha_l}(q) = 4\pi\int dr r^2 K_l(q,r) S^l_{\alpha_1 \ldots \alpha_l}(q),
  \label{momkpeqn}
\end{equation}
which relates the correlation moments $R^l_{\alpha_1 \ldots \alpha_l}(q)$ to source 
moments $S^l_{\alpha_1 \ldots \alpha_l}(q)$. Eq.~\ref{momkpeqn} is similar to the 
1D Koonin-Pratt equation (cf. Eq. \ref{kpeqn}) but pertains to moments describing 
different ranks of angular anisotropy $l$. 

	The mathematical structure of Eq.\ref{momkpeqn} is similar to that of 
Eq.~\ref{kpeqn}. Therefore, the same 1D Imaging technique (describe earlier) can be used to 
invert each correlation moment $R^l_{\alpha_1 \ldots \alpha_l}(q)$ to extract the 
corresponding source moment $S^l_{\alpha_1 \ldots \alpha_l}(q)$. Subsequently, 
the total 3D source function is calculated by combining the source moments 
for each $l$ as in equation (\ref{eqn4})
\begin{equation}
 S(\mathbf{r}) = \sum_l \sum_{\alpha_1 \ldots \alpha_l}
   S^l_{\alpha_1 \ldots \alpha_l}(r) \,A^l_{\alpha_1 \ldots \alpha_l} (\Omega_\mathbf{r}).
\label{eqn5}
\end{equation}
%
%
%
\begin{figure}[htb]
\begin{center}
\includegraphics*[angle=0, width=6.5cm]{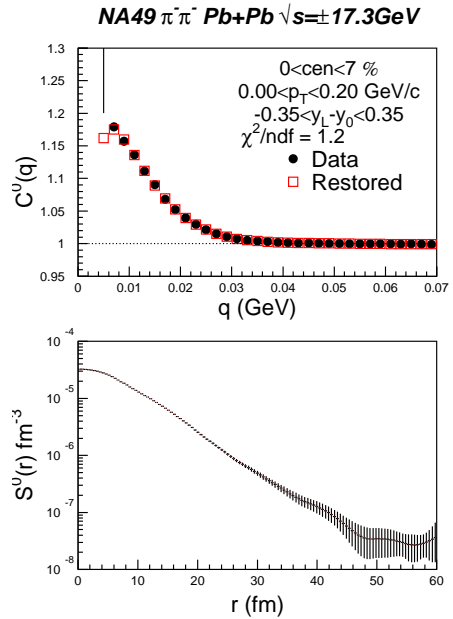}
\end{center}
\vskip-0.75cm
\caption{\emph{ \small l=0 moment $C^0(q) = R^0(q)+1$ (top panel) and imaged 
source moment $S^0(r)$ (bottom panel) for mid-rapidity low $p_T$ $\pi^-\pi^-$ pairs from  
central Pb+Pb collisions ($\sqrt s = 17.3$ GeV). The open squares in the top panel 
represent the correlation moment restored from the imaged source moment $S^0(r)$
\label{160gevs0}
}}
\end{figure}
\begin{figure}[htb]
\begin{center}
\includegraphics*[angle=0, width=6.5cm]{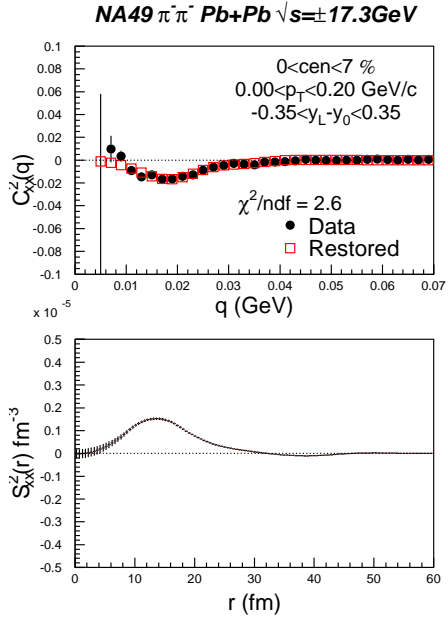}
\end{center}
\vskip-0.75cm
\caption{\emph{ \small Same as Fig.~\ref{160gevs0} but for the l=2 moment $C^2_{xx}(q) = R^2_{xx}(q)$.  
Open squares in top panel represent correlation moment restored from imaged source 
moment $S^2_{xx}(r)$.
\label{160gevsx2}
}}
\end{figure}

	Figure \ref{160gevs0} shows the imaged $l=0$ moment $S^0$ for the
input l=0 moment $C^0(q) = R^0(q)+1$ for mid-rapidity low $p_T$ $\pi^-\pi^-$ pairs 
from central Pb+Pb collisions (bottom panel). This source image looks essentially 
exponential-like. Inserting the extracted image into the 1D Koonin-Pratt 
equation (Eq.~\ref{kpeqn}) yields the restored moment shown as open squares in 
the top panel. The good agreement between the input and restored
correlation moments serves as a consistency check of the imaging procedure.

\begin{figure}[htb]
\begin{center}
\includegraphics*[angle=0, width=6.5cm]{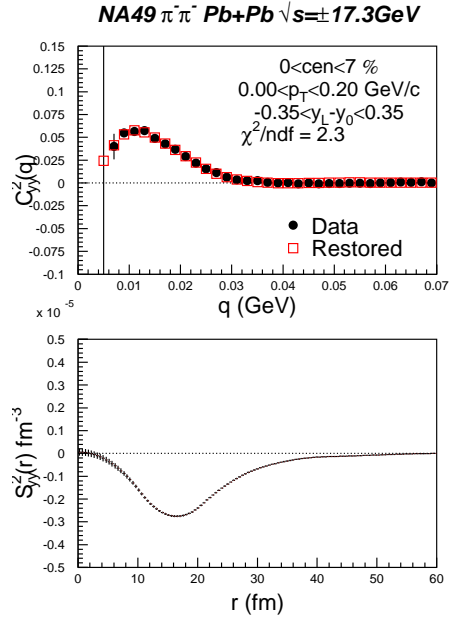}
\end{center}
\vskip-0.75cm
\caption{\emph{ \small Same as Fig.~\ref{160gevsx2} but for the l=2 moment $C^2_{yy}(q) = R^2_{yy}(q)$. 
Open squares in top panel represent correlation moment restored from imaged source 
moment $S^2_{yy}(r)$.
\label{160gevsy2}
}}
\end{figure}

	The sources images for the $l=2$ moments $R^2_{xx}$ and $R^2_{yy}$ are 
shown in Figs.~\ref{160gevsx2} and \ref{160gevsy2}. They show that the anti-correlation 
exhibited by $R^2_{xx}$ results in a positive $l=2$ source moment contribution 
$S^2_{xx}$. That is, the negative contribution of $R^2_{xx}$ plus 
$R^0$ causes the overall correlation function to be narrower in the $x$ direction 
compared to the angle-average correlation function. A narrower correlation function 
in $x$ gives a broader source function in $x$; hence, the positive contribution 
of $S^2_{xx}$ to $S^0$. 
%
%

	Figure \ref{160gevsy2} shows an opposite effect for the image for the $R^2_{yy}$ 
correlation moment. Here, the positive contribution of this correlation moment results 
in a negative contribution of the $l=2$ source moment $S^2_{yy}$. The positive 
contribution of $R^2_{yy}$ plus $R^0$ causes the overall correlation function to 
be broader in the $y$ direction compared to the angle-average correlation function. 
The resulting narrower source function (in $y$) reflects the negative contribution 
of $S^2_{yy}$ to $S^0$. 

	As discussed earlier, the imaged moments are combined to give the resulting 
pair separation distributions in the out-, side- and long-directions.
%
%

\subsection{Extracted source functions}

\begin{figure}[htb]
\begin{center}
\includegraphics*[angle=0, width=7cm]{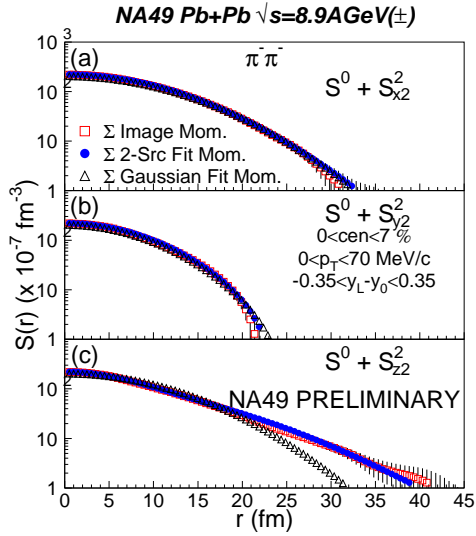}
\end{center}
\vskip-1.2cm
\caption{\emph{ \small Total source function in (a) x (out) (b) y (side) (c) z (long) 
direction for low $p_T$ mid-rapidity $\pi^-\pi^-$ pairs from central
Pb+Pb collisions at $\sqrt s = 8.9$ GeV. 
Source image and fits are indicated. 
\label{40gevsrc}
}}
\end{figure}

\begin{figure}[htb]
\begin{center}
\includegraphics*[angle=0, width=7cm]{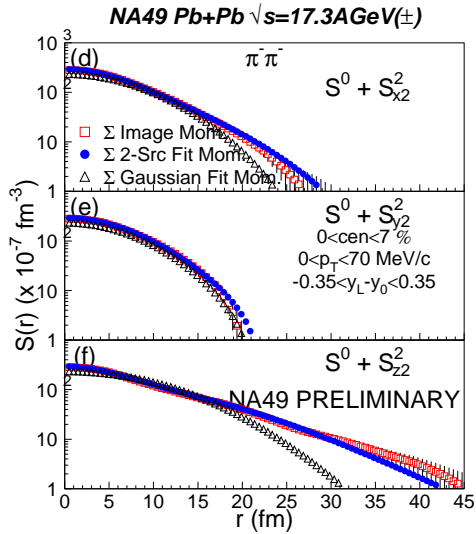}
\end{center}
\vskip-1.2cm
\caption{\emph{ \small Same as Fig.\ref{40gevsrc} but for $\sqrt s = 17.3$ GeV. 
}}
\label{160gevsrc}
\end{figure}

\begin{figure}[htb]
\begin{center}
\includegraphics*[angle=0, width=7cm]{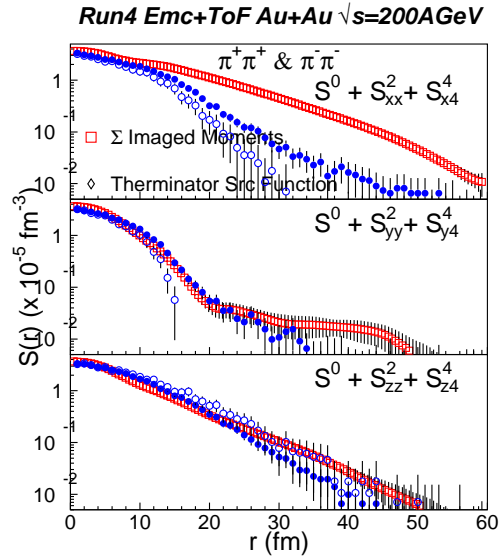}
\end{center}
\caption{\emph{ \small Source images for the x (out) (a) y (side) (b) and 
z (long) (c) directions for $.2 < p_T < .35$ GeV/c mid-rapidity $\pi^-\pi^-$ pairs 
from central Au+Au collisions at $\sqrt s = 200$ GeV. Model calculations 
from the Therminator code are also indicated.
\label{AuAu_src_roots200}
}}
\end{figure}

	As discussed in the previous section,
%
\[
 S(\mathbf{r}) = \sum_l \sum_{\alpha_1 \ldots \alpha_l}
   S^l_{\alpha_1 \ldots \alpha_l}(r) \,A^l_{\alpha_1 \ldots \alpha_l} (\Omega_\mathbf{r}).
\]
Therefore, the net source functions in the $x$ (out-), $y$ (side-) and $z$ (long-) 
directions are simply given by the sum of the 1D source $S^0$ and the 
corresponding higher $l$ contributions $S^2_{ii}$ where i= x, y, z. 
Figs. \ref{40gevsrc} - \ref{AuAu_src_roots200} show the source functions 
extracted for Pb+Pb and Au+Au collisions for the collision energies indicated.
For the Pb+Pb collisions, source moments were found to be significant only for  
the multi-polarities $l=0$ and $l=2$ for $0 < p_T < 70$ MeV/c. 
For Au+Au ($.2 < p_T < .36$ GeV/c) significant source moments were also 
obtained for $l=4$. 

	In Figs. \ref{40gevsrc}  and  \ref{160gevsrc} the results 
obtained from source imaging and fits to the correlation moments 
are shown. Clear indications for non-Gaussian contributions to these 
source functions are apparent. A detailed analysis of these source functions 
is still ongoing. However, a number of observations can be drawn. First, the 
ratio of the RMS radii of the source functions in the $x$ and $y$ directions 
are 1.3$\pm$0.1 and and 1.2$\pm$0.1 respectively. 
This deviation from unity points to a finite pion emission time. Here, it is 
important to emphasize that the rather low selection $0 < p_T < 70$ MeV/c, ensures
that the Lorentz factor $\gamma$ is essentially 1. That is, any extension in the 
$x$ direction due to kinematic transformation from the LCMS to the PCMS is 
negligible. 
%
%

	Another important observation is that the RMS pair separation 
in the $z$ direction (cf. Figs. \ref{40gevsrc}(c) and  \ref{160gevsrc}(f)) are 11~fm  
and 12~fm respectively. These dimensions are much larger than the Lorentz-contracted 
nuclear diameters of 3~fm at $\sqrt s = 8.9$ GeV and 1.5~fm at $\sqrt s = 17.3$ GeV. 
In fact, they can be taken as the RMS pair separation resulting from the longitudinal 
spread of nuclear matter created by the passage of the two nuclei. Since the latter
are moving with almost the speed of light, one can infer a lower bound formation 
time for the created nuclear matter. This is estimated to be 
$\sim 8$~fm/c at $\sqrt s = 8.9$ GeV and $\sim 10$~fm/c at $\sqrt s = 17.3$ GeV.

	Figure \ref{AuAu_src_roots200} show representative source images extracted for 
Au+Au collisions at $\sqrt s = 200$ GeV. For comparison, source functions obtained 
from initial calculations performed with the Therminator simulation 
code \cite{Kisiel:2006is} are also plotted. The filled and 
open circles show the results from calculations {\em with} and {\em without} resonance 
emissions turned on. Fig. \ref{AuAu_src_roots200}b shows that the calculations
which include resonance emissions can account for an apparent tail in 
the distribution for the $y$ (side) direction. However, Fig. \ref{AuAu_src_roots200}a
suggest that such resonance contributions are not sufficient to account for an 
apparent extension in the $x$ (out) direction. An extensive effort to map out 
the source functions for different models and model parameters is currently 
underway. It is expected that a comparison between the results of such calculations 
and data will provide tighter constraints for the space-time evolution for 
collision systems spanning the AGS - SPS - RHIC bombarding energy range.

\section{Conclusions}

	In summary, we have presented and discussed several techniques for femtoscopic 
measurements spanning collision energies from the AGS to RHIC. These methods of analysis 
provide a robust technique for detailed investigations of the reaction dynamics. Initial 
results from an ambitious program to measure a ``full" femetoscopic excitation function,
indicate that the emission source functions are decidedly non-Gaussian for a 
broad range of collision energies, and particle emission times can be reliably measured.


%
\bibliography{Lacey_ISMD06_Proc} 
%

%
%
%
%

\end{document}